\pdfoutput=1
\documentclass[12pt,a4paper]{article}

\usepackage{ifthen} 
\newboolean{pdflatex}
\setboolean{pdflatex}{true} 

\newboolean{articletitles}
\setboolean{articletitles}{true} 

\newboolean{uprightparticles}
\setboolean{uprightparticles}{false} 

\def\paperauthors{LHCb collaboration} 
\def\paperasciititle{Observation of two new excited Xib0 states decaying to Lambdab K-pi+} 
\def\papertitle{Observation of two new excited 

$\Xibz$ states decaying to $\Lb K^-\pip$} 
\def\paperkeywords{{High Energy Physics}, {LHCb}} 
\def\papercopyright{\the\year\ CERN for the benefit of the LHCb collaboration} 
\def\paperlicence{CC BY 4.0 licence}
\def\paperlicenceurl{https://creativecommons.org/licenses/by/4.0/}

\def\lbkpi{\ensuremath{\Lb\Km\pip}\xspace}

\def\xiblow{\ensuremath{\PXi_b(6327)^0}\xspace}
\def\xibhigh{\ensuremath{\PXi_b(6333)^0}\xspace}

\def\lblcpi{\ensuremath{\Lb\to\Lc\pim}\xspace}
\def\lblcpipipi{\ensuremath{\Lb\to\Lc\pim\pip\pim}\xspace}

\def\lcpi{\ensuremath{\Lc\pim}\xspace}
\def\lcpipipi{\ensuremath{\Lc\pim\pip\pim}\xspace}

\usepackage[top=1in, bottom=1.25in, left=1in, right=1in]{geometry}

\usepackage{microtype}
\usepackage{lineno}  
\usepackage{xspace} 
\usepackage{caption} 

\usepackage{graphicx}  
\usepackage{color}
\usepackage{colortbl}
\graphicspath{{./figs/}} 

\usepackage{amsmath} 
\usepackage{amssymb}
\usepackage{amsfonts}
\usepackage{upgreek} 

\newcommand*\patchAmsMathEnvironmentForLineno[1]{%
\expandafter\let\csname old#1\expandafter\endcsname\csname #1\endcsname
\expandafter\let\csname oldend#1\expandafter\endcsname\csname
end#1\endcsname
 \renewenvironment{#1}%
   {\linenomath\csname old#1\endcsname}%
   {\csname oldend#1\endcsname\endlinenomath}%
}
\newcommand*\patchBothAmsMathEnvironmentsForLineno[1]{%
  \patchAmsMathEnvironmentForLineno{#1}%
  \patchAmsMathEnvironmentForLineno{#1*}%
}
\AtBeginDocument{%
\patchBothAmsMathEnvironmentsForLineno{equation}%
\patchBothAmsMathEnvironmentsForLineno{align}%
\patchBothAmsMathEnvironmentsForLineno{flalign}%
\patchBothAmsMathEnvironmentsForLineno{alignat}%
\patchBothAmsMathEnvironmentsForLineno{gather}%
\patchBothAmsMathEnvironmentsForLineno{multline}%
\patchBothAmsMathEnvironmentsForLineno{eqnarray}%
}

\usepackage{hyperxmp}

\usepackage[pdftex,
            pdfauthor={\paperauthors},
            pdftitle={\paperasciititle},
            pdfkeywords={\paperkeywords},
            pdfcopyright={Copyright (C) \papercopyright},
            pdflicenseurl={\paperlicenceurl}]{hyperref}

\usepackage[colorinlistoftodos,textsize=scriptsize]{todonotes}

\usepackage[bottom,flushmargin,hang,multiple]{footmisc}

\usepackage[all]{hypcap} 

\usepackage{xspace} 
\usepackage{upgreek}

\def\lhcb   {\mbox{LHCb}\xspace}

\def\cms    {\mbox{CMS}\xspace}

\def\MagUp {\mbox{\em Mag\kern -0.05em Up}\xspace}

\ifthenelse{\boolean{uprightparticles}}%
{

 \def\Pmu         {\ensuremath{\upmu}\xspace}

 \def\Ppi         {\ensuremath{\uppi}\xspace}

 \def\Ppsi        {\ensuremath{\uppsi}\xspace}

 \def\PDelta      {\ensuremath{\Delta}\xspace}                 
 \def\PXi         {\ensuremath{\Xi}\xspace}                 
 \def\PLambda     {\ensuremath{\Lambda}\xspace}                 
 \def\PSigma      {\ensuremath{\Sigma}\xspace}                 
 \def\POmega      {\ensuremath{\Omega}\xspace}                 
 \def\PUpsilon    {\ensuremath{\Upsilon}\xspace}

 \def\PB      {\ensuremath{\mathrm{B}}\xspace}                 
                  
 \def\PD      {\ensuremath{\mathrm{D}}\xspace}

 \def\PJ      {\ensuremath{\mathrm{J}}\xspace}                 
 \def\PK      {\ensuremath{\mathrm{K}}\xspace}

 \def\Pb      {\ensuremath{\mathrm{b}}\xspace}                 
 \def\Pc      {\ensuremath{\mathrm{c}}\xspace}                 
 \def\Pd      {\ensuremath{\mathrm{d}}\xspace}

 \def\Pi      {\ensuremath{\mathrm{i}}\xspace}

 \def\Pp      {\ensuremath{\mathrm{p}}\xspace}                 
 \def\Pq      {\ensuremath{\mathrm{q}}\xspace}                 
                  
 \def\Ps      {\ensuremath{\mathrm{s}}\xspace}                 
                  
 \def\Pu      {\ensuremath{\mathrm{u}}\xspace}

 \def\thebaroffset{0.0em}
}
{

 \def\Pmu         {\ensuremath{\mu}\xspace}

 \def\Ppi         {\ensuremath{\pi}\xspace}

 \def\Ppsi        {\ensuremath{\psi}\xspace}                 
                  
 \mathchardef\PDelta="7101
 \mathchardef\PXi="7104
 \mathchardef\PLambda="7103
 \mathchardef\PSigma="7106
 \mathchardef\POmega="710A
 \mathchardef\PUpsilon="7107
                  
 \def\PB      {\ensuremath{B}\xspace}                 
                  
 \def\PD      {\ensuremath{D}\xspace}

 \def\PJ      {\ensuremath{J}\xspace}                 
 \def\PK      {\ensuremath{K}\xspace}

 \def\Pb      {\ensuremath{b}\xspace}                 
 \def\Pc      {\ensuremath{c}\xspace}                 
 \def\Pd      {\ensuremath{d}\xspace}

 \def\Pi      {\ensuremath{i}\xspace}

 \def\Pp      {\ensuremath{p}\xspace}                 
 \def\Pq      {\ensuremath{q}\xspace}                 
                  
 \def\Ps      {\ensuremath{s}\xspace}                 
                  
 \def\Pu      {\ensuremath{u}\xspace}

 \def\thebaroffset{0.18em}
}
\newcommand{\offsetoverline}[2][\thebaroffset]{\kern #1\overline{\kern -#1 #2}}%

\makeatletter
\ifcase \@ptsize \relax
  \newcommand{\miniscule}{\@setfontsize\miniscule{4}{5}}
\or
  \newcommand{\miniscule}{\@setfontsize\miniscule{5}{6}}
\or
  \newcommand{\miniscule}{\@setfontsize\miniscule{5}{6}}
\fi
\makeatother

\DeclareRobustCommand{\optbar}[1]{\shortstack{{\miniscule (\rule[.5ex]{1.25em}{.18mm})}
  \\ [-.7ex] $#1$}}

\def\mumu       {{\ensuremath{\Pmu^+\Pmu^-}}\xspace}

\def\quark     {{\ensuremath{\Pq}}\xspace}

\def\uquark    {{\ensuremath{\Pu}}\xspace}

\def\dquark    {{\ensuremath{\Pd}}\xspace}

\def\squark    {{\ensuremath{\Ps}}\xspace}

\def\cquark    {{\ensuremath{\Pc}}\xspace}

\def\bquark    {{\ensuremath{\Pb}}\xspace}

\def\pion   {{\ensuremath{\Ppi}}\xspace}

\def\pip    {{\ensuremath{\pion^+}}\xspace}
\def\pim    {{\ensuremath{\pion^-}}\xspace}

\def\kaon    {{\ensuremath{\PK}}\xspace}

\def\KorKbar {\kern \thebaroffset\optbar{\kern -\thebaroffset \PK}{}\xspace}

\def\Kp      {{\ensuremath{\kaon^+}}\xspace}
\def\Km      {{\ensuremath{\kaon^-}}\xspace}

\def\D       {{\ensuremath{\PD}}\xspace}

\def\DorDbar {\kern \thebaroffset\optbar{\kern -\thebaroffset \PD}\xspace}

\def\Dp      {{\ensuremath{\D^+}}\xspace}
\def\Dm      {{\ensuremath{\D^-}}\xspace}

\def\DpDm    {\ensuremath{\Dp {\kern -0.16em \Dm}}\xspace}

\def\B       {{\ensuremath{\PB}}\xspace}

\def\BorBbar {\kern \thebaroffset\optbar{\kern -\thebaroffset \PB}\xspace}

\def\Bd      {{\ensuremath{\B^0}}\xspace}

\def\BdorBdbar {\kern \thebaroffset\optbar{\kern -\thebaroffset \Bd}\xspace}
\def\Bu      {{\ensuremath{\B^+}}\xspace}

\def\Bs      {{\ensuremath{\B^0_\squark}}\xspace}

\def\BsorBsbar {\kern \thebaroffset\optbar{\kern -\thebaroffset \Bs}\xspace}

\def\jpsi     {{\ensuremath{{\PJ\mskip -3mu/\mskip -2mu\Ppsi}}}\xspace}

\def\Y#1S{\ensuremath{\PUpsilon{(#1S)}}\xspace}

\def\proton      {{\ensuremath{\Pp}}\xspace}

\def\Lz          {{\ensuremath{\PLambda}}\xspace}

\def\LorLbar     {\kern \thebaroffset\optbar{\kern -\thebaroffset \PLambda}\xspace}

\def\Sigmares    {{\ensuremath{\PSigma}}\xspace}

\def\Xires       {{\ensuremath{\PXi}}\xspace}

\def\Lc          {{\ensuremath{\Lz^+_\cquark}}\xspace}

\def\Lb           {{\ensuremath{\Lz^0_\bquark}}\xspace}

\def\Sigmab       {{\ensuremath{\Sigmares_\bquark}}\xspace}
\def\Sigmabp      {{\ensuremath{\Sigmares_\bquark^+}}\xspace}

\def\Xib          {{\ensuremath{\Xires_\bquark}}\xspace}
\def\Xibz         {{\ensuremath{\Xires^0_\bquark}}\xspace}

\newcommand{\decay}[2]{\ensuremath{#1\!\to #2}\xspace} 

\def\to                 {\ensuremath{\rightarrow}\xspace}

\def\AT#1     {\ensuremath{A_{\mathrm{T}}^{#1}}\xspace}           

\def\C#1      {\ensuremath{\mathcal{C}_{#1}}\xspace}                       
\def\Cp#1     {\ensuremath{\mathcal{C}_{#1}^{'}}\xspace}                    
\def\Ceff#1   {\ensuremath{\mathcal{C}_{#1}^{\mathrm{(eff)}}}\xspace}        
\def\Cpeff#1  {\ensuremath{\mathcal{C}_{#1}^{'\mathrm{(eff)}}}\xspace}       
\def\Ope#1    {\ensuremath{\mathcal{O}_{#1}}\xspace}                       
\def\Opep#1   {\ensuremath{\mathcal{O}_{#1}^{'}}\xspace}                    

\newcommand{\aunit}[1]{\ensuremath{\text{\,#1}}}       

\newcommand{\tev}{\aunit{Te\kern -0.1em V}\xspace}
\newcommand{\gev}{\aunit{Ge\kern -0.1em V}\xspace}
\newcommand{\mev}{\aunit{Me\kern -0.1em V}\xspace}
\newcommand{\kev}{\aunit{ke\kern -0.1em V}\xspace}
\newcommand{\ev}{\aunit{e\kern -0.1em V}\xspace}
 
\newcommand{\mevc}{\ensuremath{\aunit{Me\kern -0.1em V\!/}c}\xspace}
\newcommand{\gevc}{\ensuremath{\aunit{Ge\kern -0.1em V\!/}c}\xspace}
\newcommand{\mevcc}{\ensuremath{\aunit{Me\kern -0.1em V\!/}c^2}\xspace}
\newcommand{\gevcc}{\ensuremath{\aunit{Ge\kern -0.1em V\!/}c^2}\xspace}

\def\fb   {\ensuremath{\aunit{fb}}\xspace}
\def\invfb   {\ensuremath{\fb^{-1}}\xspace}

\def\ps   {\ensuremath{\aunit{ps}}\xspace}

\newcommand{\chisq}{\ensuremath{\chi^2}\xspace}

\newcommand{\chisqip}{\ensuremath{\chi^2_{\text{IP}}}\xspace}

\newcommand{\chisqvtx}{\ensuremath{\chi^2_{\text{vtx}}}\xspace}
\newcommand{\chisqvtxndf}{\ensuremath{\chi^2_{\text{vtx}}/\mathrm{ndf}}\xspace}
\newcommand{\chisqtrkndf}{\ensuremath{\chi^2_{\text{trk}}/\mathrm{ndf}}\xspace}

\def\gsim{{~\raise.15em\hbox{$>$}\kern-.85em
          \lower.35em\hbox{$\sim$}~}\xspace}
\def\lsim{{~\raise.15em\hbox{$<$}\kern-.85em
          \lower.35em\hbox{$\sim$}~}\xspace}

\def\sqs   {\ensuremath{\protect\sqrt{s}}\xspace}

\def\pt         {\ensuremath{p_{\mathrm{T}}}\xspace}

\def\ptot       {\ensuremath{p}\xspace}

\def\evtgen     {\mbox{\textsc{EvtGen}}\xspace}

\def\geant      {\mbox{\textsc{Geant4}}\xspace}

\def\photos     {\mbox{\textsc{Photos}}\xspace}

\def\pythia     {\mbox{\textsc{Pythia}}\xspace}

\def\tell1  {TELL1\xspace}
\def\ukl1   {UKL1\xspace}

\usepackage{cite} 
\usepackage{mciteplus}

\usepackage{longtable} 

\begin{document}

\renewcommand{\thefootnote}{\fnsymbol{footnote}}
\setcounter{footnote}{1}


\begin{titlepage}
	\pagenumbering{roman}

	\vspace*{-1.5cm}
	\centerline{\large EUROPEAN ORGANIZATION FOR NUCLEAR RESEARCH (CERN)}
	\vspace*{1.5cm}
	\noindent
	\begin{tabular*}{\linewidth}{lc@{\extracolsep{\fill}}r@{\extracolsep{0pt}}}
		\ifthenelse{\boolean{pdflatex}}
		{\vspace*{-1.5cm}\mbox{\!\!\!\includegraphics[width=.14\textwidth]{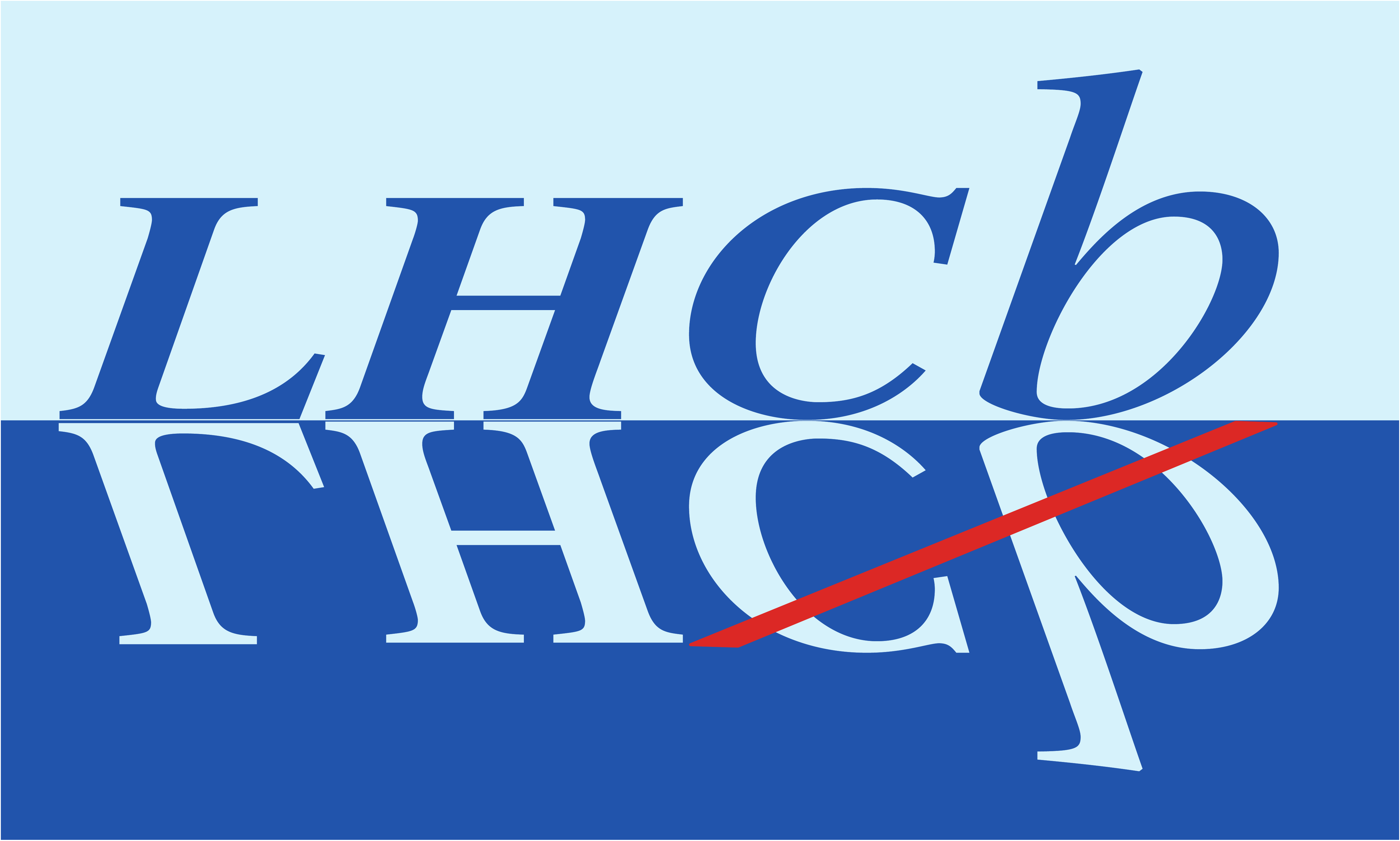}} & &}%
		{\vspace*{-1.2cm}\mbox{\!\!\!\includegraphics[width=.12\textwidth]{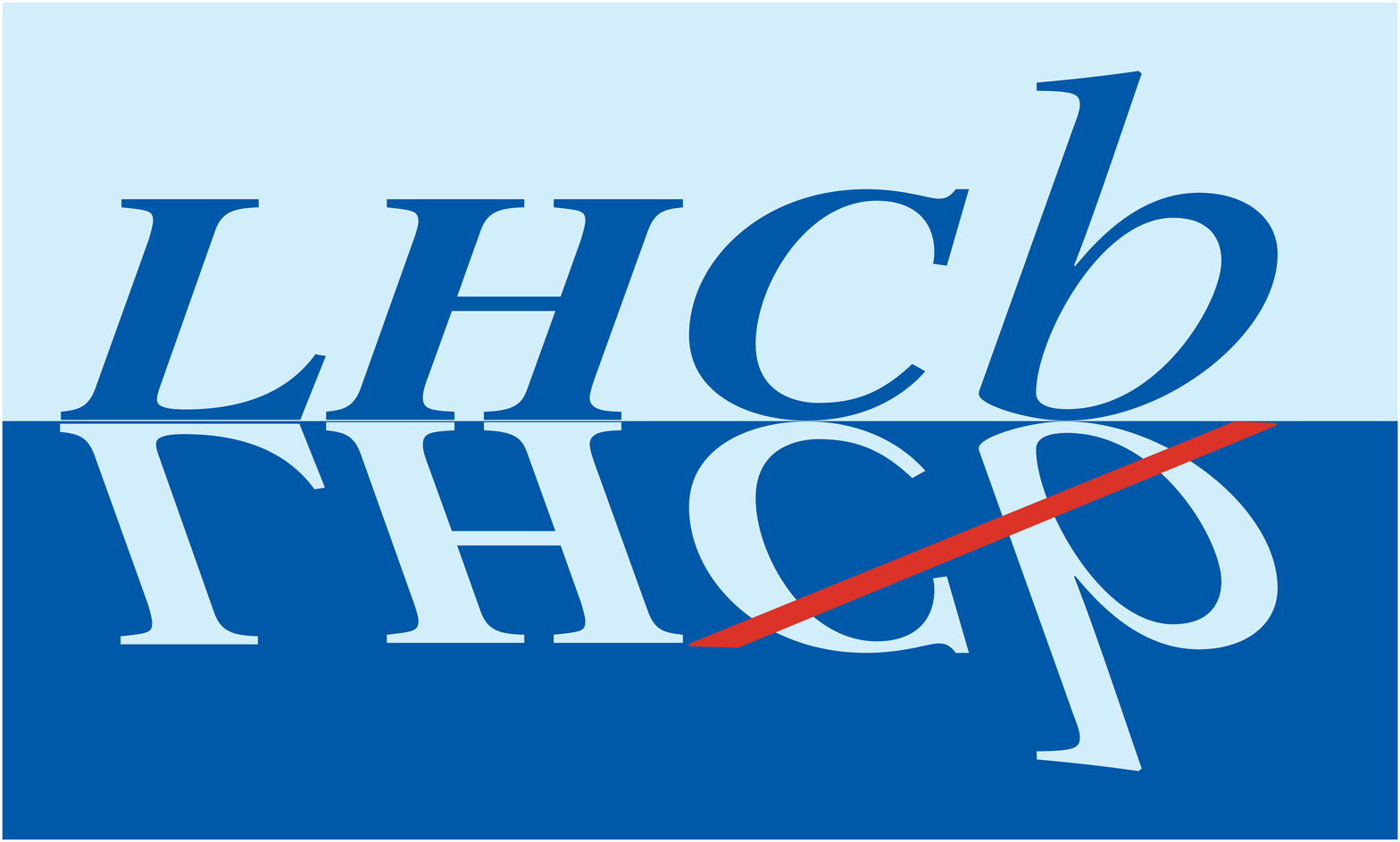}} & &}%
		\\
		& & CERN-EP-2021-188 \\  
		& & LHCb-PAPER-2021-025 \\  
		& & \today \\ 
	\end{tabular*}

	\vspace*{4.0cm}

	{\normalfont\bfseries\boldmath\huge
		\begin{center}
			\papertitle 
		\end{center}
	}

	\vspace*{2.0cm}

	\begin{center}
		\paperauthors\footnote{Authors are listed at the end of this Letter.}
	\end{center}

	\vspace{\fill}

	\begin{abstract}
		\noindent
		Two narrow resonant states are observed in the $\lbkpi$ mass spectrum using a data sample of proton-proton collisions at a center-of-mass energy of 13\tev, collected by the LHCb experiment 
		and corresponding to an integrated luminosity of $6\invfb$. 
		The minimal quark content of the $\lbkpi$ system indicates that these are excited $\Xibz$ baryons. 
		The masses of the \xiblow and \xibhigh states are 
			\mbox{$m(\xiblow) = 6327.28 ^{\,+\,0.23}_{\,-\,0.21} \pm 0.12 \pm 0.24\mev$}
			and
			\mbox{$m(\xibhigh) = 6332.69 ^{\,+\,0.17}_{\,-\,0.18} \pm 0.03 \pm 0.22\mev$},
		respectively, with a mass splitting of \mbox{$\Delta m = 5.41 ^{\,+\,0.26}_{\,-\,0.27} \pm 0.12 \mev$},
		where the uncertainties are statistical, systematic and  due to the \Lb mass measurement. The measured natural widths of these states are consistent with zero, with upper limits of $\Gamma(\xiblow)<2.20~(2.56)\mev$ and  $\Gamma(\xibhigh)<1.60~(1.92)\mev$ at a 90\%~(95\%) credibility level. 
		The significance of the two-peak hypothesis is larger than nine (five) Gaussian standard deviations compared to the no-peak (one-peak) hypothesis.
		The masses, widths and resonant structure of the new states are in good agreement with the expectations for a doublet of 1$D$ $\Xibz$ resonances. 
	\end{abstract}

	\vspace*{1.0cm}

	\begin{center}
		Published in
		Phys.~Rev.~Lett.
		128 (2022) 162001
	\end{center}

	\vspace{\fill}

	{\footnotesize 
	\centerline{\copyright~\papercopyright. \href{\paperlicenceurl}{\paperlicence}.}}
	\vspace*{2mm}

\end{titlepage}


\newpage
\setcounter{page}{2}
\mbox{~}
%
%
%
%


\renewcommand{\thefootnote}{\arabic{footnote}}
\setcounter{footnote}{0}

\cleardoublepage


\pagestyle{plain} 
\setcounter{page}{1}
\pagenumbering{arabic}




In the constituent quark model \cite{GellMann:1964nj,Zweig:352337}, baryons comprising a $\bquark$ quark and two light quarks ($\bquark\quark\quark^{\prime}$) form multiplets based on the symmetries of their flavor, spin, and spatial wave functions \cite{Klempt:2009pi}.
If $q$ and $q'$ are $\uquark$ or $\dquark$ quarks, these beauty baryons are classified into the $\Lb$ and $\Sigmab$ baryon families, where the light diquark spin $j_{\quark\quark^{\prime}}$ is $0$ and $1$, respectively.
A beauty baryon containing one $\squark$ quark ($\bquark\squark\quark$) forms the $\Xib$ or $\PXi_b^{\prime}$ family depending on whether the light diquark spin $j_{\squark\quark}$ is $0$ or $1$.
Most of the ground states of these beauty-baryon families have been observed \cite{PDG2020}. 
Beyond that many radially and orbitally excited states with higher masses are predicted by theory \cite{Chen:2019ywy,Chen:2018orb,Chen:2018vuc,Ebert:2011kk,Yao:2018jmc,Ebert:2007nw,Roberts:2007ni,Chen:2014nyo,Wang:2017kfr,Kawakami:2019hpp}. 
In recent years, several excited $\Lb$ states have been observed  \cite{LHCB-PAPER-2012-012,LHCb-PAPER-2019-025,LHCb-PAPER-2019-045}. 
This motivates further investigations of the lesser known excited $\Xib$ states, as 
the $\Lb$ and $\Xib$ states have similar properties due to the approximate $SU(3)$ flavor symmetry \cite{Chen:2019ywy}. 
Recently, the \lhcb collaboration reported the observation of the $\PXi_b(6227)^-$ baryon~\cite{LHCb-PAPER-2018-013} and its isospin partner, the $\PXi_b(6227)^0$ baryon~\cite{LHCb-PAPER-2020-032}, and the 
 \cms collaboration reported the observation of the $\PXi_b(6100)^-$ baryon~\cite{CMS:2021rvl}. 
No other excited $\Xib$ states have been observed.  
There are predictions of two 1$D$ $\Xibz$ baryons mainly decaying through the $\PSigma_b^{(*)}\kaon$ and $\PXi_{b}^{*,\prime}\pion$ modes \cite{Chen:2019ywy,Yao:2018jmc}, where the label 1$D$ refers to a unity radial quantum number and a $D$ wave (orbital momentum $L=2$) between the $\bquark$ quark and the light diquark system. A search for these predicted excited states and studies of their masses, widths and decay patterns can provide valuable validation to the understanding of quantum chromodynamics, which features the behavior of strong interactions.  
The $\PSigma_b^{(*)}\kaon$ mode results in a \lbkpi final state. 

In this Letter, the observation of a structure with two narrow peaks in the $\lbkpi$ mass spectrum is presented (the inclusion of charge-conjugated processes is implied throughout this Letter), using proton-proton ($\proton\proton$) collision data collected by the $\lhcb$ experiment at a center-of-mass energy of $\sqs=13\tev$
, corresponding to an integrated luminosity of $6\invfb$. 
A measurement of the mass and width of each state, and an investigation of the resonant structure contributing to the three-body  decays of the excited $\Xibz$ states  are performed. 
The resulting properties are consistent with those predicted states for a 1$D$ \Xibz doublet~\cite{Chen:2019ywy,Yao:2018jmc}, hereafter referred to as $\xiblow$ and $\xibhigh$ states.


The \lhcb detector \cite{Alves:2008zz,LHCb-DP-2014-002} is a single-arm forward spectrometer covering the \mbox{pseudorapidity} range $2<\eta<5$, designed for the study of particles containing $\bquark$ or $\cquark$ quarks. 
The detector includes a high\nobreakdash-precision tracking system consisting of a silicon\nobreakdash-strip vertex detector surrounding the~$\proton\proton$~interaction region~\cite{LHCb-DP-2014-001}, a~large-area silicon-strip detector located upstream of a dipole magnet with a bending power of about $4{\mathrm{\,Tm}}$, and three stations of silicon-strip detectors and straw drift tubes \cite{LHCb-DP-2013-003} placed downstream of the magnet.
The tracking system provides a measurement of the~momentum, \ptot, of charged particles with a relative uncertainty that varies from 0.5\% at low momentum to 1.0\% at 200\gev (natural units with $c=\hbar=1$ are used throughout this Letter).
The momentum scale of the tracking system is calibrated using samples of $\decay{\jpsi}{\mumu}$ and $\decay{\Bu}{\jpsi\Kp}$ decays collected concurrently with the data sample used for this analysis \cite{LHCb-PAPER-2012-048,LHCb-PAPER-2013-011}.
Different types of charged hadrons are distinguished using information from two ring-imaging Cherenkov detectors~\cite{LHCb-DP-2012-003}.
Photons, electrons and hadrons are identified by a calorimeter system consisting of
scintillating-pad and preshower detectors, an electromagnetic and a hadronic calorimeter. 
Muons are identified by a system composed of alternating layers of iron and multiwire proportional chambers~\cite{LHCb-DP-2012-002}.
The online event selection is performed by a trigger \cite{LHCb-DP-2012-004, LHCb:2018zdd} which consists of a~hardware stage, based on information from the calorimeter and muon systems, followed by a software stage, in which charged particles are reconstructed and a real-time analysis is performed. At the hardware stage, the \proton\proton collision events are required to have a muon with high \pt or a hadron, photon or electron with large transverse energy deposited in the calorimeter. The software trigger requires a two-, three- or four-track secondary vertex with a significant displacement from any primary \proton\proton collision vertex (PV), and at least one charged particle with a large transverse momentum and inconsistent with originating from any PV. 
Simulation is required to model the effects of the detector acceptance, the imposed selection requirements and the detector resolution on the invariant mass spectrum.
The $pp$ collisions are generated using \pythia~\cite{Sjostrand:2007gs} with a specific \lhcb configuration~\cite{LHCb-PROC-2010-056}.
Decays of unstable particles are described by \evtgen~\cite{Lange:2001uf}, using \photos~\cite{davidson2015photos} and by the \geant toolkit~\cite{Allison:2006ve, *Agostinelli:2002hh,LHCb-PROC-2011-006}.


The $\Lb$ baryon is reconstructed using its decays into the $\lcpi$ and $\lcpipipi$ final states, where the \Lc baryon subsequently decays to the \proton\Km\pip final state. 
All charged final\nobreakdash-state particles are required to have particle-identification information consistent with their respective mass hypotheses.
A neural network is used to reject misreconstructed tracks~\cite{DeCian:2255039}. 
To suppress combinatorial background from the PV, 
the final-state protons, kaons and pions are required to have transverse momenta $\pt>100\mev$, $p>1\gev$ and $\chisqip > 4$ with respect to all PVs in the event, where \chisqip of a particle is the difference in \chisq of the vertex fit of a given PV, with the particle being included or excluded. 
The reconstructed \Lc vertex is required to have $\chisqvtxndf < 10$ and $\chisq_{\rm FD} > 36$, where $\chisqvtx$ is the \chisq value of the vertex fit per degree of freedom, and $\chisq_{\rm FD}$ is the $\chisq$ distance from the closest PV. 
The reconstructed mass must be within a window of $\pm 25$\mev ($\pm18\mev$) of the known \Lc mass \cite{PDG2020} for $\lcpi$ ($\lcpipipi$) candidates.
The tighter mass cut applied in the $\lcpipipi$ sample is due to its higher background level. 
The selected $\Lc$ candidates are further combined with pion candidates to form \Lb candidates,  
where the \Lb candidates are required to have $\chisqvtxndf < 10$ and a reconstructed proper lifetime larger than $0.2\ps$.


A boosted decision tree (BDT) algorithm \cite{Roe:2004na,AdaBoost,Hocker:2007ht} is used to enhance further the signal purity of the \Lb samples.
The choice of the training variables follows similar strategies as that for \lblcpi~\cite{LHCb-PAPER-2019-025} and \lblcpipipi~\cite{LHCb-PAPER-2020-032} analysis. For both modes, the following variables are used: 
the $\chisqip$ values, the $\pt$, and the $\chisqvtxndf$ of \Lc and \Lb candidates, the $\chisq_{\rm FD}$, the angle between the reconstructed momentum and flight direction of the $\Lc$ and $\Lb$ candidates, and the quality of particle identification for final-state pions, kaons and protons.  
In addition, the $\chisqip$ value and \pt of the pion originating from the \Lb decay are used for the \lblcpi mode, while the $\chisqip$ and flight-distance significance of the pion from the \Lb decay, the vertex-fit quality and the invariant mass of the $\pim\pip\pim$ system are used for the  \lblcpipipi mode. 
The BDT classifier is trained on data using background-subtracted \cite{Pivk:2004ty} \Lb candidates to represent the signal sample, and \Lb candidates 
with $\lcpi$ and $\lcpipipi$ invariant mass ranging between 5700 and 5800\mev (higher-mass sideband) to represent the background sample. 
Since the training samples are also used for the further analysis, the $k$-fold cross-validation technique \cite{kFold} with $k=10$ is applied to avoid any possible effect of overtraining. 
The chosen working point of the BDT classifier rejects half of the combinatorial background, with a negligible reduction on the signal efficiency. 
The resulting \Lb signal yields in the selected samples are $966\,000$ and $533\,000$ for the \lblcpi and \lblcpipipi decays, respectively.
The invariant mass of selected \lblcpi and \lblcpipipi candidates is shown in the Supplemental Material of this Letter.

The selected \Lb candidates are further combined with a kaon and pion stemming from the $\proton\proton$ interaction point to form the \Lb\Km\pip candidates. 
The pion and kaon candidates  are required to have $\chisqip<9$,  $\chisqtrkndf<3$    
and to have $\ptot>1500\mev,~\pt(\kaon)>800\mev,~\pt(\pion)>250\mev$, where $\chisqtrkndf$ is the track fit $\chisq$ per degree of freedom. 
Then \Lb, pion and kaon are combined to form \Lb\kaon\pion candidates, which are required to have a vertex-fit $\chisq$ smaller than 20 and an invariant mass of the ${\Lb\pion}$ system smaller than 5850\mev. 
This is $10\mev$ higher than the predicted kinematic maximum value of the \Lb\pion mass for the \xiblow and \xibhigh states~\cite{Chen:2019ywy}. 
The \lbkpi combinations containing the signal decays are hereafter referred to as the right-sign (RS) sample.
For a better modeling of the background shape from random combinations of $\Lb$, $\Kp$ and $\pim$ candidates, the wrong-sign (WS) candidates are reconstructed in the \Lb\Kp\pim final state. 
The same selections are applied to both the RS and WS samples. 
To improve the mass resolution of the excited $\Xibz$ candidates, the reconstructed mass is redefined as  $m(\Lb\kaon\pion)\equiv M(\Lb\kaon\pion) - M(\Lc\pim (\pip \pim))+M_{\Lb}$, where  $M_{\Lb}$ is the known \Lb mass measured by the \lhcb collaboration~\cite{LHCB-PAPER-2017-011}, 
$M(\Lb\kaon\pion)$, $M(\Lc\pim)$ and $M(\Lc\pim\pip\pim)$ are the invariant masses calculated constraining~\cite{Hulsbergen:2005pu} the \Lc mass to the world average value~\cite{PDG2020}, and that the \Lb\kaon\pion and \Lb candidates originate in the PV.

\begin{figure}[t]
    \centering
    \includegraphics[width=0.45\textwidth]{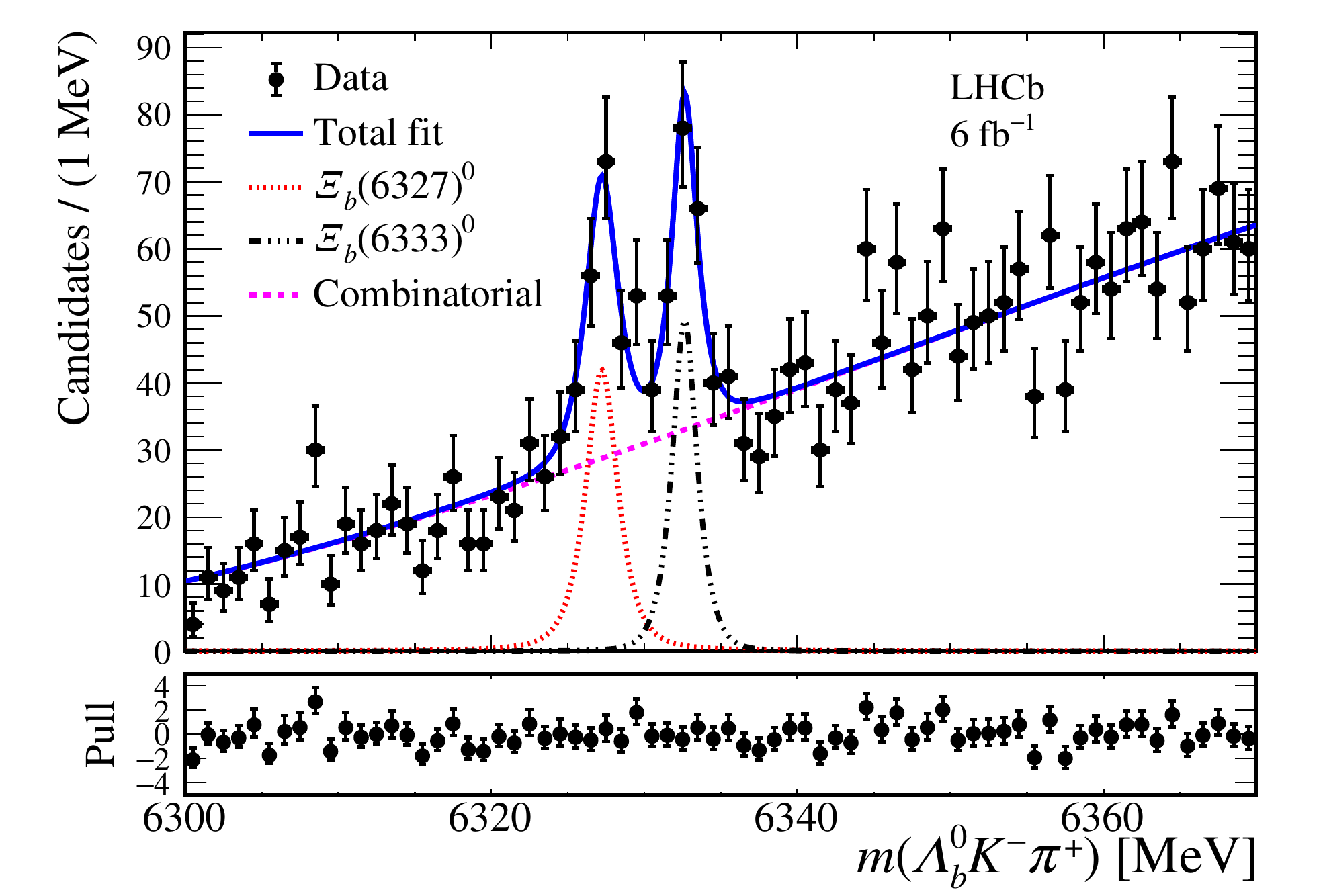}
    \includegraphics[width=0.45\textwidth]{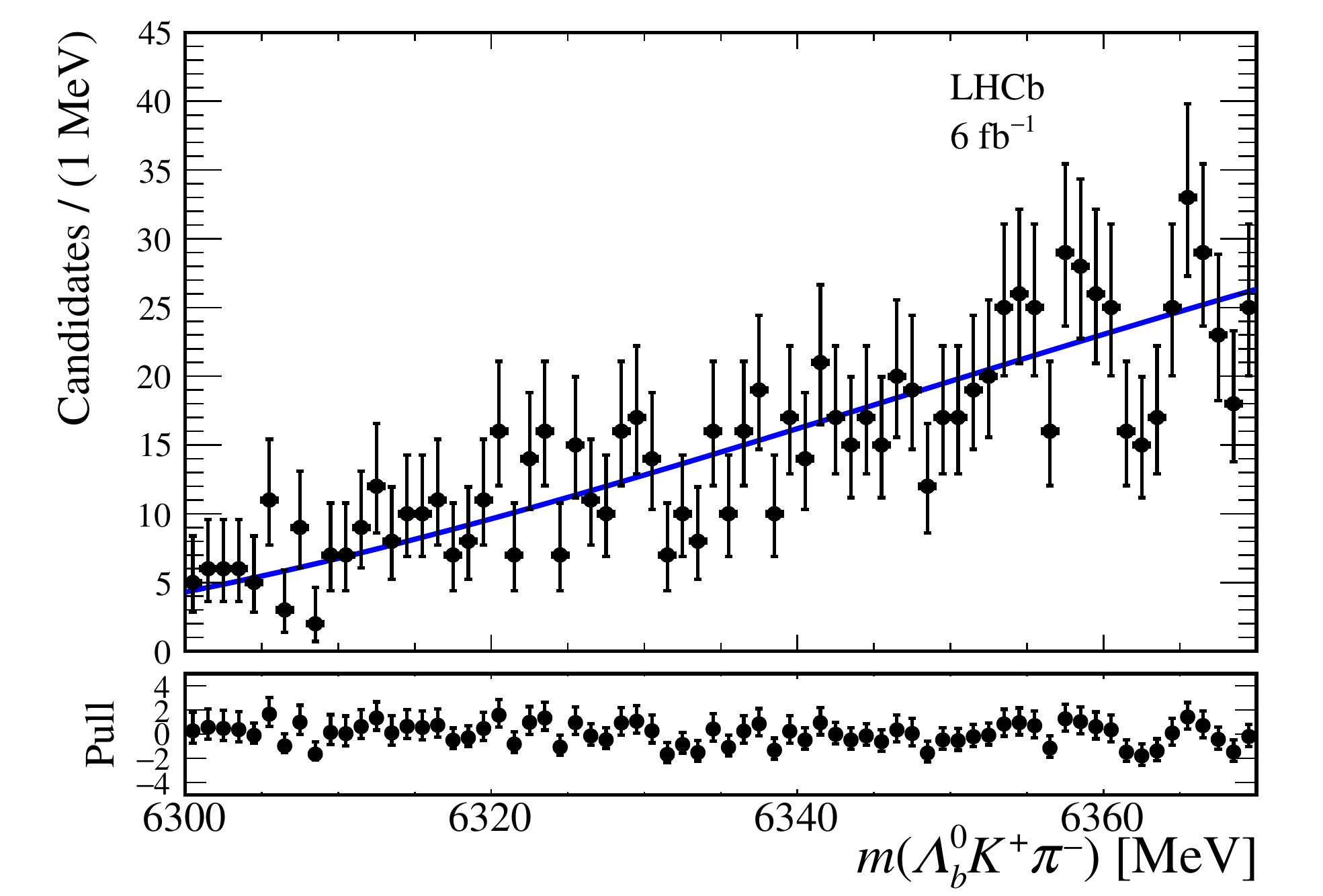}
    \caption{Invariant mass distributions of \Lb\kaon\pion candidates from (left) RS and (right) WS samples. The fit projections are overlaid. The black points with error bars correspond to the data, and the blue line shows the total fit projection. Individual fit components are listed in the legend. }
    \label{fig:xib_mass}
\end{figure}

The $m(\Lb\kaon\pion)$ distributions of the RS and WS samples are shown in Fig.~\ref{fig:xib_mass}.
Two narrow peaks can be seen around $6330\mev$ in the \Lb\Km\pip mass spectrum, while no significant peaking structure is visible in the \Lb\Kp\pim system. 
A simultaneous extended unbinned maximum-likelihood fit is performed to the RS and WS samples to determine the properties of the peaking structure. 
Each peak in the RS sample is modeled as a constant\nobreakdash-width relativistic Breit--Wigner (RBW) function \cite{Jackson:1964zd} defined as 
\begin{align*}
f_{\mathrm{sig}}(m(\Lb\kaon\pion))=\frac{C_{\mathrm{sig}}}{(m^2-m^2(\Lb\kaon\pion))^2 +m^2\Gamma^2},
\end{align*}
where $C_{\mathrm{sig}}$ is a normalization factor, 
$m$ is the mass of \Xibz state, and $\Gamma$ is its mass\nobreakdash-independent width.
The RBW function is convolved with a resolution model, parameterized as a symmetric variant of the Apollonios function \cite{Santos:2013gra}. 
The parameters of the resolution function are fixed to values determined from simulation.
The background component, which is present in both the RS and WS samples, is described by a threshold function \mbox{$f_{\mathrm{bkg}}(m(\Lb\kaon\pion))= C_{\mathrm{bkg}}(m(\Lb\kaon\pion)-m_t)^{a_0}e^{-a_{1}(m(\Lb\kaon\pion)-m_t)}$,  where $C_{\mathrm{bkg}}$} is a normalization factor, $a_0$ and $a_1$ are free parameters in the fit, $m_t$ is the minimum mass of the \lbkpi combination, which corresponds to the sum of the \Lb~\cite{LHCB-PAPER-2017-011}, pion and kaon~\cite{PDG2020} masses. 
The same shape parameters for the background, $a_0$ and $a_1$, are shared by the RS and WS samples.

The masses and widths of the \xiblow and \xibhigh states are measured to be
\begin{align*}
    m(\xiblow)&= 6327.28_{\,-\,0.21}^{\,+\,0.23}\mev,\\
    m(\xibhigh)&= 6332.69_{\,-\,0.18}^{\,+\,0.17}\mev,\\
    \Gamma(\xiblow)&=0.93_{\,-\,0.60}^{\,+\,0.74}\mev,\\
    \Gamma(\xibhigh)&=0.25_{\,-\,0.25}^{\,+\,0.58}\mev,
\end{align*}
with a mass splitting of $\Delta m \equiv m(\xibhigh) - m(\xiblow) =  5.41 ^{\,+\,0.26}_{\,-\,0.27} \mev$, and the resulting \xiblow and \xibhigh signal yields are $134\pm27$ and $117\pm24$, respectively.  The uncertainties listed above are statistical only.

A likelihood-ratio test statistic is used to estimate the global significance of the two observed states. 
For a first estimation based on Wilks’ theorem \cite{Wilks:1938dza}, it is assumed that without these peaks, the value of twice the change of log-likelihood $\Delta_{2\log\mathcal{L}} \equiv 2\log (\mathcal{L}_{\rm max}/\mathcal{L}_0)$ follows a $\chisq$ distribution. 
The symbol $\mathcal{L}_{\rm max}$ indicates the maximum likelihood value with both peaks included in the fit model, while $\mathcal{L}_0$ is the value obtained from a null hypothesis with no peak or one peak included.
The number of degrees of freedom of the $\chisq$ distribution is the difference of the number of floating parameters in the default fit and under the null hypothesis. 
With this method, the significance of the two-peak hypothesis is $10.4\sigma$ and $6.6\sigma$ with respect to the no-peak and one-peak hypotheses, respectively, where $\sigma$ represents a Gaussian standard deviation. 
Pseudoexperiments are performed 
as an alternative method for estimating the statistical significance of the two-peak hypothesis with respect to the null hypothesis, including the no-peak and one-peak assumptions.  
A total of 200\,000 pseudoexperiments are performed based on the null hypothesis and the $\Delta_{2\log\mathcal{L}}$ value is estimated for each of these. The $\Delta_{2\log\mathcal{L}}$ distribution is parameterized as a shape of which the tail can be modeled using a $\chi^2$ distribution, with the number of degrees of freedom allowed to take non-integer values and determined by fitting the $\Delta_{2\log\mathcal{L}}$ distribution of the pseudoexperiments. When performing the pseudoexperiments, the look-elsewhere effect is considered by constraining the peaking position in several different mass intervals which combined together cover the full mass interval shown in Fig.~\ref{fig:xib_mass}. The $p$ value of the two-peak hypothesis is estimated to be $10.2\sigma$ and $6.6\sigma$, with no-peak and one-peak assumptions set as the null hypothesis, respectively. 
The significance from pseudoexperiments for the two hypotheses is consistent with the values from the Wilks’ theorem \cite{Wilks:1938dza}, and the lowest values are taken as the statistical significance of the two observed states.

\begin{figure}[t]
    \centering
    \includegraphics[width=0.45\textwidth]{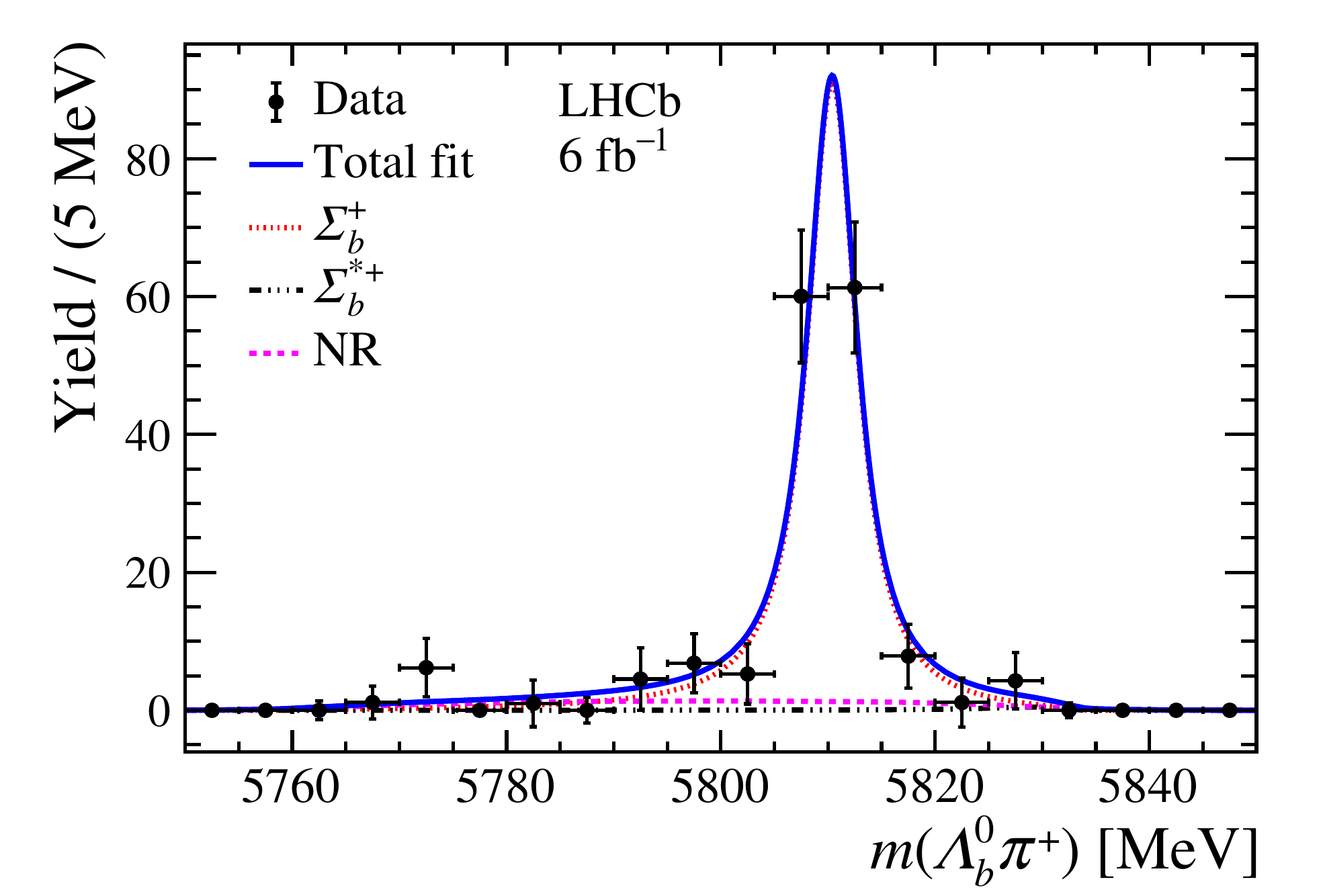}
    \includegraphics[width=0.45\textwidth]{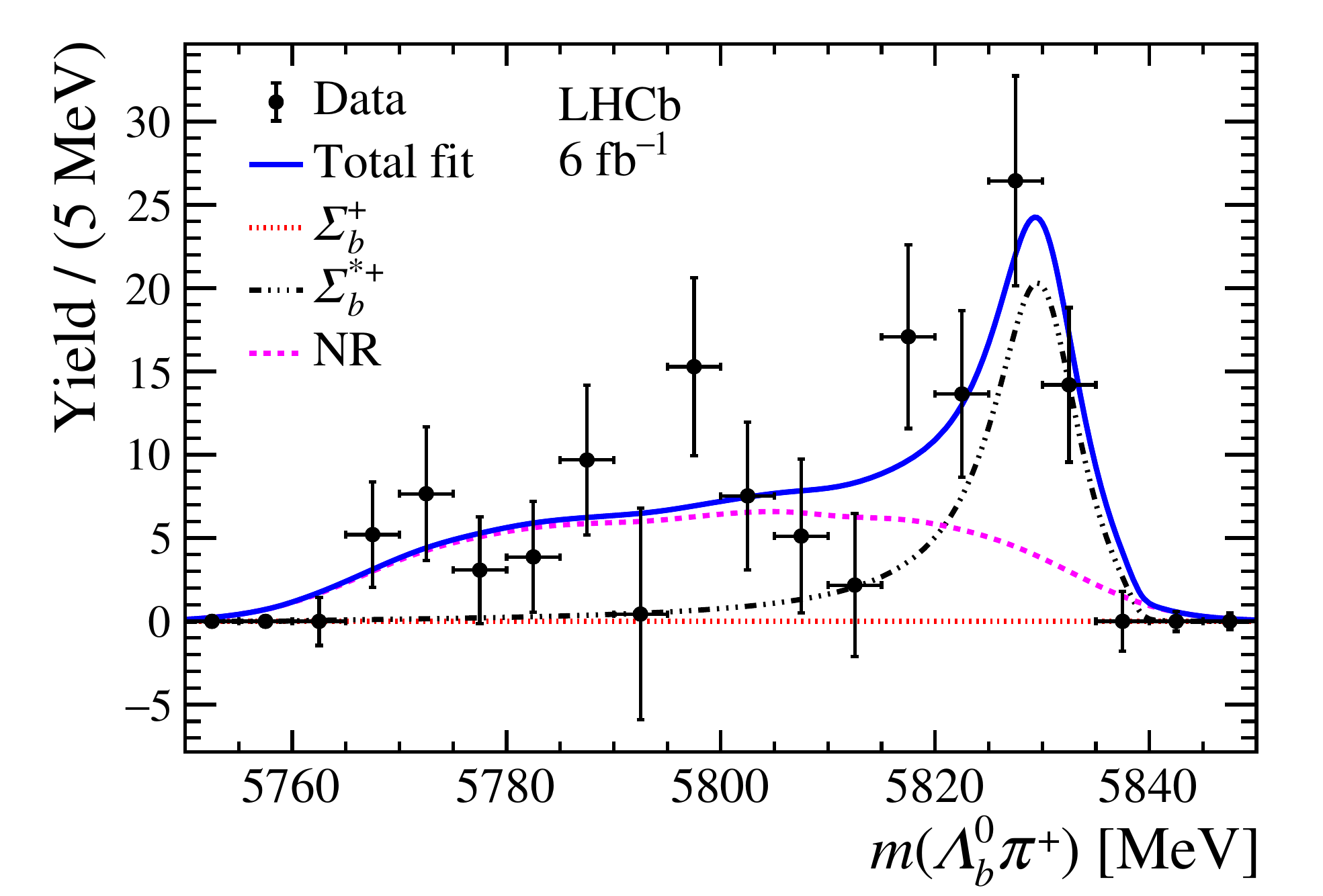}
    \caption{Signal yields of the (left) \xiblow and (right) \xibhigh states determined by mass fits to data samples in 5\mev slices of the \Lb\pion mass spectrum. The black points with error bars correspond to the yields of the \xiblow or \xibhigh states, the blue solid lines are the total fit projections. Each individual component of the fit model is indicated in the legend.}
    \label{fig:xib_resonance}
\end{figure}

To study the resonant structure in the excited \Xibz decays, several \Lb\Km\pip mass fits to data samples in 5\mev wide slices of the \Lb\pion mass regions are performed. 
The default fit model described previously is used, 
with the masses and widths of the two \Xibz states fixed to the default values.
The signal yields of \xiblow and \xibhigh states as a  function of the \Lb\pion mass are shown in Fig.~\ref{fig:xib_resonance}, where significant peaking structure corresponding to the $\Sigmabp$ or $\PSigma_b^{*+}$ states can be seen. 
The projections of the binned maximum-likelihood fits to the distributions are overlaid. 
The $\Sigmabp$ and $\PSigma_b^{*+}$ contributions are modeled using a RBW amplitude with a mass-dependent width~\cite{Jackson:1964zd,Blatt:1952ije}, with the phase-space density of a three-body decay~\cite{PDG2020} and Blatt--Weisskopf barrier factors~\cite{Blatt:1952ije} considered when constructing the fit function.
The non-resonant contribution (NR) is modeled using uniform-phase-space simulation and obtained by a kernel estimation~\cite{Cranmer:2000du}.
The interference between the NR component and the $\PSigma_b^{(*)+}$ resonances is not considered. 
The \xiblow state predominantly decays to \Sigmabp\Km.
About half of the \xibhigh baryons decay without $\Lb\pip$ resonances, while the rest is dominated by the decay through the $\PSigma_b^{*+}$ intermediate structure. 
The resonant structure is consistent with the theoretical predictions for a doublet of 1$D$ \Xibz states \cite{Chen:2019ywy,Yao:2018jmc}, where the \Sigmabp\Km process dominates the decay of the lighter state, while the $\PSigma_b^{*+}\Km$ mode has a significant contribution to the decay of the heavier one.

Several sources of systematic uncertainties are considered for the mass and width measurements.
The uncertainty related to the momentum scale is evaluated by varying the momentum scale within its known uncertainty of $3\times 10^{-4}$ \cite{LHCb-PAPER-2012-048,LHCb-PAPER-2013-011}, and  determining the effect on the mass and width parameters. 
To estimate the systematic uncertainties related to the choice of the functions used to model the signal and background shapes of the \Lb\kaon\pion invariant mass spectra, several alternative fit models are used. 
The RBW functions with mass-dependent widths~\cite{Jackson:1964zd,Blatt:1952ije} are used to model the \Xibz states, where the phase-space factors and barrier factors are calculated 
assuming the \Xibz decays to occur through the \Xibz\to\Sigmabp\Km or $\Xibz\to\PSigma_b^{*+}\Km$ two-body processes. 
For these RBW functions, the  orbital angular momentum of the \Sigmabp\Km system is varied between 0 and 3 and the  Blatt--Weisskopf barrier radius~\cite{Blatt:1952ije} between 1.0 and $5.0\gev^{-1}$.  
Polynomial functions with order between 2 and 4 are used as alternative models to describe the background and to  estimate the corresponding systematic uncertainties. An alternative \Lb\kaon\pion invariant-mass fit is performed using the default model but with only the RS sample, and the variation of the mass and width parameters are considered as the systematic uncertainties accounting for the potential discrepancy between the shapes of WS sample and RS background components. 
Alternative resolution functions, either the sum of two Crystal Ball~\cite{Skwarnicki:1986xj} or of two Gaussian functions, are used to model the detector resolution effect.
To consider the potential difference of mass resolution between data and simulation, the resolution determined by simulation is varied by $\pm10\%$~\cite{LHCB-PAPER-2017-036,LHCb-PAPER-2018-013,LHCb-PAPER-2019-025,LHCB-PAPER-2014-061,LHCb-PAPER-2019-005} and the impact on the fit result is assigned as uncertainty. 
About 10\% of the selected $pp$ collision events contain more than one $\Lb\Km\pip$ candidates, and these are retained in the data sample. 
As an alternative event-selection algorithm, only one candidate is kept for each event, and the variations of the mass and width parameters are treated as an additional source of systematic uncertainty. 
The significance of the two-peak structure is estimated based on all alternative fit models, and the smallest value is taken as the significance including systematic uncertainties. 
The significance of the two-peak structure is $9.9\sigma$ and $5.8\sigma$ with respect to the no-peak and one-peak hypotheses, respectively. 
As the reconstructed mass of the \Xibz candidates is defined using $M_{\Lb}$ as an input, a corresponding uncertainty is considered. 
The value of $M_{\Lb}$ is taken from the \Lb mass measurement performed by the \lhcb collaboration~\cite{LHCB-PAPER-2017-011}, where the \Lb candidates are reconstructed with several \Lb decays excluding the \mbox{\Lb\to\Lc\pim} and \mbox{\Lb\to\Lc\pim\pip\pim} in this analysis. Therefore, the statistical uncertainty of the \Lb mass measurement is treated as an uncorrelated source of uncertainty. 
Among all sources of uncertainties of the \Lb mass result, only the systematic uncertainty related to the momentum scale is fully correlated with the corresponding uncertainty in this analysis, whereas the other systematic uncertainties are assumed to be uncorrelated.
The total systematic uncertainty on the mass and width is calculated as the sum in quadrature of the different sources and summarized in Table~\ref{tab:sys}. 

The method to set an upper limit on the width is based on the Bayesian credibility level with a flat prior for non-negative width~\cite{PDG2020,Narsky:1999kt}.
The upper limits of the widths of the \xiblow and \xibhigh states are evaluated by convolving the likelihood profiles with the total uncertainty of the width parameters in Table~\ref{tab:sys}, and finding the values that cover 90\% or 95\% of the integrated probability.

\begin{table}[t]
	\caption{Systematic uncertainties on masses and widths (in ${\rm MeV}$) for the \xiblow and \xibhigh states, and uncertainties from the $\Lb$ mass measurement~\cite{LHCB-PAPER-2017-011}. The systematic uncertainties due to the imperfect knowledge of the momentum scale (syst. momentum scale), systematic uncertainties from other sources (syst. excl. momentum scale), and the statistical uncertainties (stat.) are also listed as individual terms in this table.} 
	\begin{center}
		\begin{tabular}{lccccc}
			\hline
			Source & \multicolumn{2}{c}{\xiblow} & \multicolumn{2}{c}{\xibhigh}\\
			& $m$ & $\Gamma$ & $m$ & $\Gamma$ & $\Delta m$\\
			\hline
			Momentum scale & 0.06 & 0.06 & 0.03 & 0.04 & 0.03\\
			Signal shape & 0.01 & 0.12 & 0.00 & 0.25 & 0.01\\
			Background shape & 0.01 & 0.17 & 0.01 & 0.15 & 0.00\\
			Resolution model & 0.05 & 0.20 & 0.01 & 0.25 & 0.05\\
			Multiple candidates & 0.09 & 0.02 & 0.01 & 0.23 & 0.11\\
			\hline
			Total systematic uncertainty & 0.12 & 0.30 & 0.03 & 0.45 & 0.12\\
			\hline
			$\Lb$ mass ({\rm syst. momentum scale}) & 0.12 & - & 0.12 & - & -\\
			$\Lb$ mass ({\rm syst. excl. momentum scale}) & 0.05 & - & 0.05 & - & -\\
			\Lb mass (stat.) & 0.16 & - & 0.16 & - & -\\
			\hline
			Total uncertainty from \Lb mass & 0.24 & - & 0.22 & - & -\\
			\hline
		\end{tabular}\\
	\end{center}
	\label{tab:sys}
\end{table}


In summary, two new states, \xiblow and \xibhigh, are observed in the \lbkpi mass spectrum, where the \Lb baryon is reconstructed in the \lcpi and \lcpipipi final states. 
The significance of the two-peak hypothesis is larger than $9\sigma$ compared to the no-peak hypothesis and $5\sigma$ compared to the one-peak hypotheses in terms of Gaussian standard deviations.
The masses of these two states are measured to be
\begin{align*}
	m(\xiblow) &= 6327.28 ^{\,+\,0.23}_{\,-\,0.21} \pm 0.12 \pm 0.24\mev,\\
	m(\xibhigh) &= 6332.69 ^{\,+\,0.17}_{\,-\,0.18} \pm 0.03 \pm 0.22\mev,
\end{align*}
where the first uncertainties are statistical, the second systematic and the third is due to the \Lb mass measurement.
The corresponding widths are consistent with zero, and upper limits at 90\% (95\%) credibility level are set,
\begin{align*}
	\Gamma(\xiblow)&<2.20~(2.56)\mev ,\\
	\Gamma(\xibhigh)&<1.60~(1.92)\mev .
\end{align*}
The mass differences between the excited \Xibz baryon and the ground state \Lb baryons are measured to be
\begin{align*}
	m(\xiblow)-M_{\Lb} = 707.66 ^{\,+\,0.23}_{\,-\,0.21} \pm 0.12\mev,\\
	m(\xibhigh)-M_{\Lb} = 713.07 ^{\,+\,0.17}_{\,-\,0.18} \pm 0.03\mev,
\end{align*}
with a mass splitting between the two \Xibz states of
\begin{align*}
\Delta m = 5.41 ^{\,+\,0.26}_{\,-\,0.27} \pm 0.12 \mev, 
\end{align*}
where the uncertainties are statistical and systematic, respectively.
This is the first observation of two states decaying to the \lbkpi final state. Their masses, widths and decay patterns are consistent with the predictions \cite{Chen:2019ywy,Yao:2018jmc} for a doublet of 1$D$ \Xibz states with $J^P=3/2^{+}$ and $5/2^{+}$.

\section*{Acknowledgements}
%
%
\noindent We express our gratitude to our colleagues in the CERN
accelerator departments for the excellent performance of the LHC. We
thank the technical and administrative staff at the LHCb
institutes.
We acknowledge support from CERN and from the national agencies:
CAPES, CNPq, FAPERJ and FINEP (Brazil); 
MOST and NSFC (China); 
CNRS/IN2P3 (France); 
BMBF, DFG and MPG (Germany); 
INFN (Italy); 
NWO (Netherlands); 
MNiSW and NCN (Poland); 
MEN/IFA (Romania); 
MSHE (Russia); 
MICINN (Spain); 
SNSF and SER (Switzerland); 
NASU (Ukraine); 
STFC (United Kingdom); 
DOE NP and NSF (USA).
We acknowledge the computing resources that are provided by CERN, IN2P3
(France), KIT and DESY (Germany), INFN (Italy), SURF (Netherlands),
PIC (Spain), GridPP (United Kingdom), RRCKI and Yandex
LLC (Russia), CSCS (Switzerland), IFIN-HH (Romania), CBPF (Brazil),
PL-GRID (Poland) and NERSC (USA).
We are indebted to the communities behind the multiple open-source
software packages on which we depend.
Individual groups or members have received support from
ARC and ARDC (Australia);
AvH Foundation (Germany);
EPLANET, Marie Sk\l{}odowska-Curie Actions and ERC (European Union);
A*MIDEX, ANR, IPhU and Labex P2IO, and R\'{e}gion Auvergne-Rh\^{o}ne-Alpes (France);
Key Research Program of Frontier Sciences of CAS, CAS PIFI, CAS CCEPP, 
Fundamental Research Funds for the Central Universities, 
and Sci. \& Tech. Program of Guangzhou (China);
RFBR, RSF and Yandex LLC (Russia);
GVA, XuntaGal and GENCAT (Spain);
the Leverhulme Trust, the Royal Society
 and UKRI (United Kingdom).

\clearpage
\section*{Appendix: Supplemental material}
\label{sec:Supplementary}


\subsection*{The \boldmath{\lblcpi} and \boldmath{\lblcpipipi} candidates}

\begin{figure}[bh]
    \centering
    \includegraphics[width=0.45\textwidth]{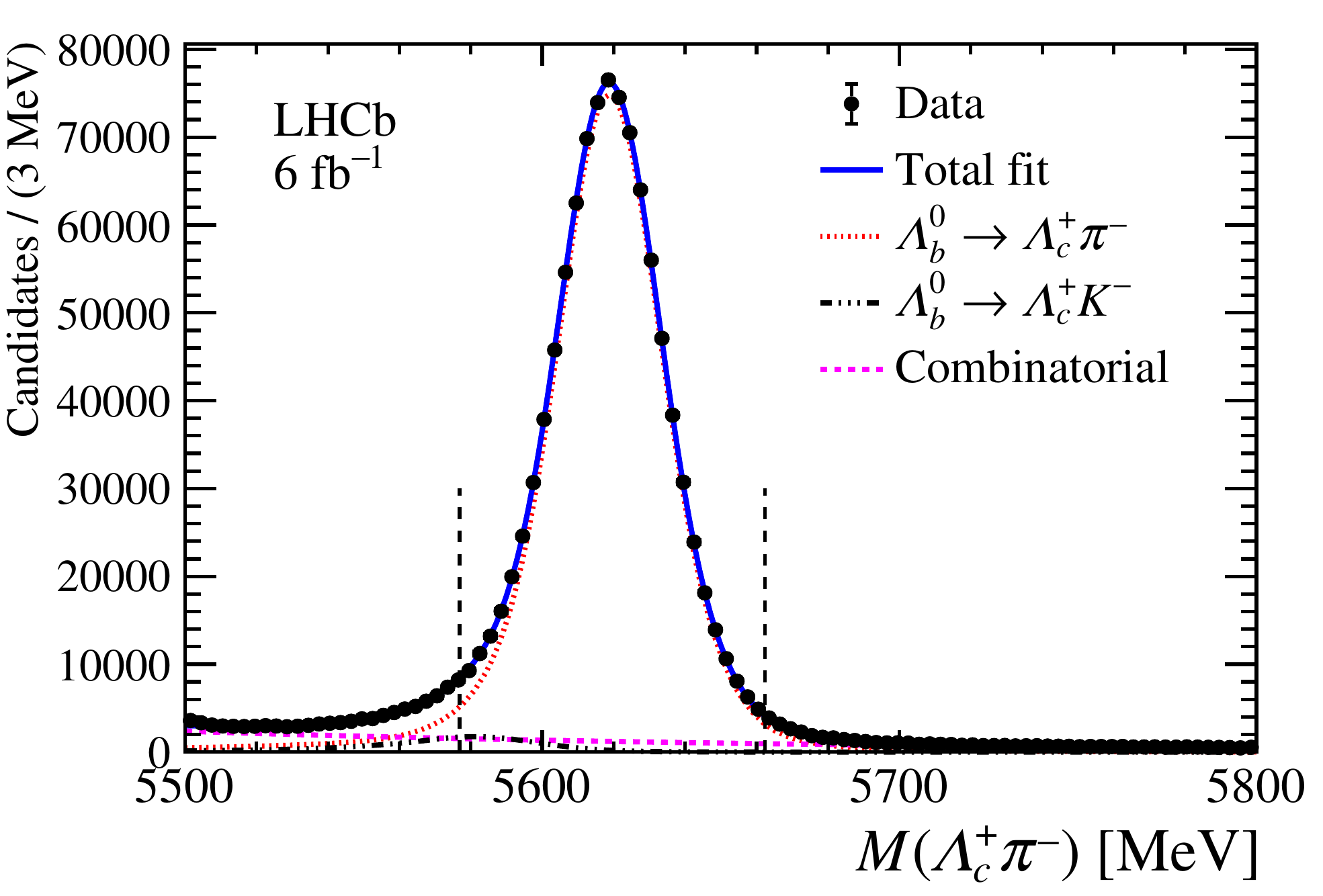}
    \includegraphics[width=0.45\textwidth]{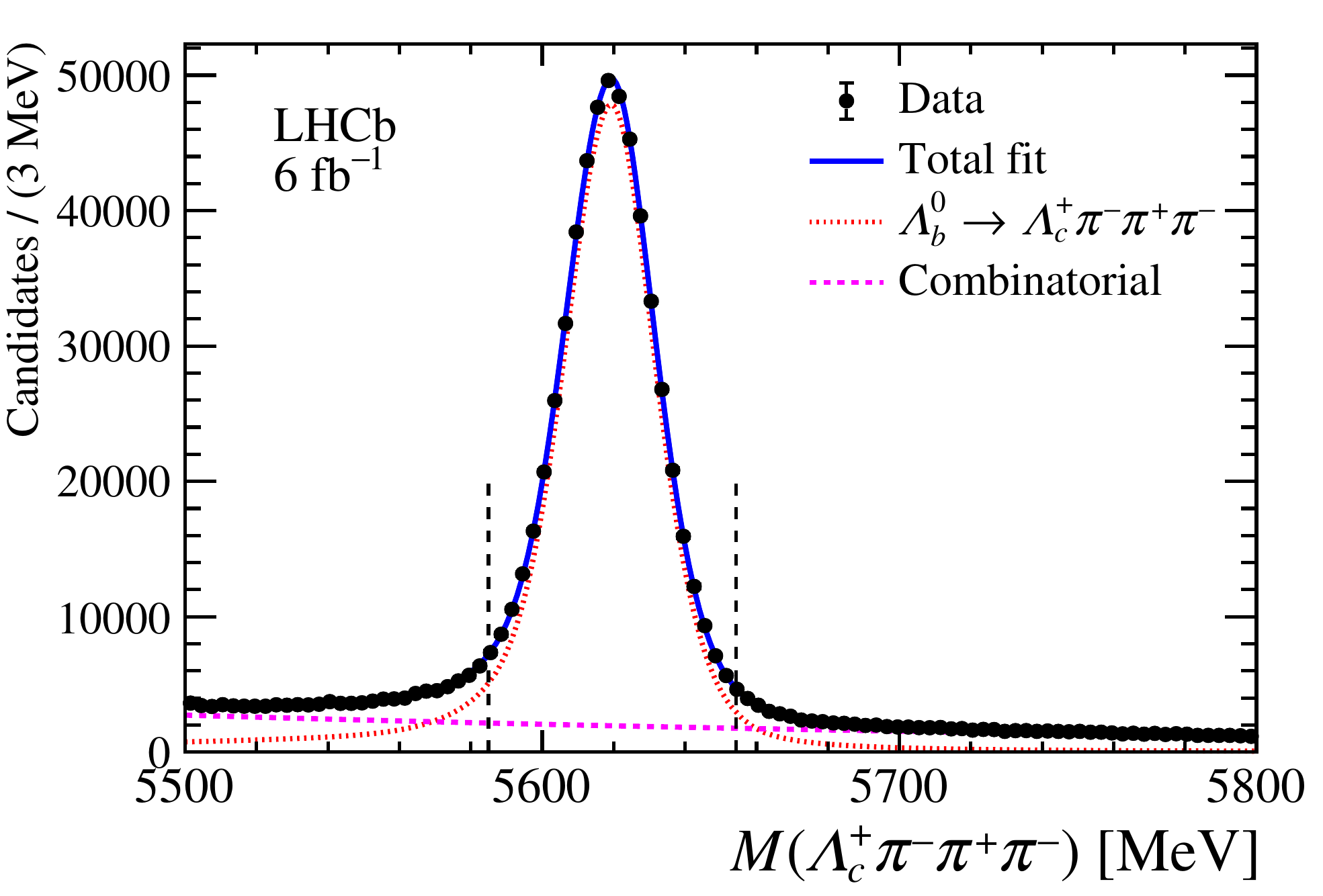}
    \caption{Invariant mass distributions of \Lb candidates reconstructed from (left) \lblcpi and (right) \lblcpipipi decays with fit projections overlaid. The black points with error bars correspond to the data, and the blue line shows the total fit projection. Individual fit components are listed in the legend. The \Lb candidates located between vertical dashed lines are used to form \Lb\kaon\pion combinations. 
    }
    \label{fig:lb_mass}
\end{figure}

The invariant mass of selected \lblcpi and \lblcpipipi candidates is shown in Fig.~\ref{fig:lb_mass}. 
To improve the \Lb-candidate mass resolution, the \Lb mass is calculated by constraining the \Lc mass to its known value \cite{PDG2020} and the \Lb baryon to originate from the associated PV \cite{Hulsbergen:2005pu}. 
Unbinned maximum-likelihood fits to the invariant mass distributions of the \Lb candidates are performed to estimate the \Lb signal yields. 
The signal component is described by the sum of a Gaussian function and two Crystal Ball functions \cite{Skwarnicki:1986xj} with the same mean value of the Gaussian cores and tail parameters determined from simulation.
For the \lblcpi mode, the background is described by 
misidentified $\Lb\to\Lc\Km$ decays, the shape of which is determined from simulation, and combinatorial background modeled by an exponential function. 
For the \lblcpipipi mode, only the combinatorial background, modeled by an exponential function, is considered. 
These fits are also shown in Fig.~\ref{fig:lb_mass}. 
The \Lb candidates are further required to have a reconstructed mass in a $2.5\sigma$  window  around the value of the \Lb mass measured by the \lhcb collaboration~\cite{LHCB-PAPER-2017-011}, where  $\sigma$ is the mass resolution of the data sample, determined to be $17.1\mev$ for \lblcpi and $13.9\mev$ for \lblcpipipi candidates.

\clearpage

\addcontentsline{toc}{section}{References}
\bibliographystyle{LHCb}
\bibliography{main,standard,LHCb-PAPER,LHCb-CONF,LHCb-DP,LHCb-TDR}

\newpage
\centerline
{\large\bf LHCb collaboration}
\begin
{flushleft}
\small
R.~Aaij$^{32}$,
A.S.W.~Abdelmotteleb$^{56}$,
C.~Abell{\'a}n~Beteta$^{50}$,
T.~Ackernley$^{60}$,
B.~Adeva$^{46}$,
M.~Adinolfi$^{54}$,
H.~Afsharnia$^{9}$,
C.~Agapopoulou$^{13}$,
C.A.~Aidala$^{87}$,
S.~Aiola$^{25}$
% ...

{\footnotesize \it

% ...
$^{9}$Universit{\'e} Clermont Auvergne, CNRS/IN2P3, LPC, Clermont-Ferrand, France\\
% ...
$^{13}$LPNHE, Sorbonne Universit{\'e}, Paris Diderot Sorbonne Paris Cit{\'e}, CNRS/IN2P3, Paris, France\\
% ...
$^{25}$INFN Sezione di Milano, Milano, Italy\\
% ...
$^{32}$Nikhef National Institute for Subatomic Physics, Amsterdam, Netherlands\\
% ...
$^{46}$Instituto Galego de F{\'\i}sica de Altas Enerx{\'\i}as (IGFAE), Universidade de Santiago de Compostela, Santiago de Compostela, Spain\\
% ...
$^{50}$Physik-Institut, Universit{\"a}t Z{\"u}rich, Z{\"u}rich, Switzerland\\
% ...
$^{54}$H.H. Wills Physics Laboratory, University of Bristol, Bristol, United Kingdom\\
% ...
$^{56}$Department of Physics, University of Warwick, Coventry, United Kingdom\\
% ...
$^{60}$Oliver Lodge Laboratory, University of Liverpool, Liverpool, United Kingdom\\
% ...
$^{68}$Syracuse University, Syracuse, NY, United States\\
% ...
$^{87}$University of Michigan, Ann Arbor, United States, associated to $^{68}$\\
\bigskip
% ...
\medskip
}
\end{flushleft}




\end{document}